# Towards generation of mJ-level ultrashort THz pulses by optical rectification

**József András Fülöp[1], László Pálfalvi[1], Matthias C Hoffmann[2] and János Hebling[1]**

[1]Department of Experimental Physics, University of Pécs, Ifjúság ú. 6, 7624 Pécs, Hungary

[2]Max Planck Research Department for Structural Dynamics, University of Hamburg, CFEL, 22607 Hamburg, Germany

E-mail: fulop@fizika.ttk.pte.hu

**Abstract.** Optical rectification of ultrashort laser pulses in $LiNbO_3$ by tilted-pulse-front excitation is a powerful way to generate near single-cycle terahertz (THz) pulses. Motivated by various applications, calculations were carried out to optimize the THz peak electric field strength. The results predict THz output with peak electric field strength on the MV/cm level in the 0.3–1.5 THz frequency range by using optimal pump pulse duration of about 500 fs, optimal crystal length, and cryogenic temperatures for reducing THz absorption in $LiNbO_3$. The THz electric field strength can be increased further to tens of MV/cm by focusing. Using optimal conditions together with the contact grating technique THz pulses with 100 MV/cm field strength and energies on the tens-of-mJ scale are feasible.

## 1. Introduction

Optical rectification (OR) of femtosecond laser pulses is an efficient method to generate ultrashort THz pulses. The highest THz pulse energies and field strengths in the few-THz frequency range were achieved by using $LiNbO_3$ (LN) as the nonlinear medium having very large effective nonlinear coefficient [1-4]. This requires tilting the pump pulse front relative to its wave front in order to satisfy the phase matching condition required for efficient THz generation [5]. Pumped by multi-mJ Ti:sapphire lasers this technique allowed to generate ultrashort THz pulses on the 10-µJ energy and 100-kV/cm electric field strength scale [1]. Such THz pulses enabled, for example, the investigation of THz-induced nonlinear optical phenomena directly in the time-domain [6] as well as time-resolved studies of ultrafast carrier dynamics in semiconductors by THz pump—THz probe measurements [7-9]. Scaling the THz energy up to 50 µJ has been recently demonstrated by using LN and the tilted-pulse-front pumping (TPFP) technique driven by medium-scale Ti:sapphire amplifier systems with pulse energies up to 120 mJ [2, 4]. Ultrashort pulses at higher THz frequencies and electric field strengths on the 10 MV/cm scale could be achieved by optical parametric amplification in GaSe [10]. However, due to the material properties of GaSe, this technique is not easily scalable.

Ultrashort THz pulses with energies in the millijoule range and electric field strengths at 100-MV/cm levels, exceeding by far what is presently available, are required by promising new applications. These include investigation of material properties and processes under the influence of extremely high quasi-static fields, multispectral single-shot imaging, particle acceleration by electromagnetic waves, and THz-assisted attosecond pulse generation [11, 12]. Some of these applications will require large-scale ultrashort-pulse laser facilities, such as ELI [13], as the pump source.

A conventional TPFP setup consists of a femtosecond pump laser, an optical grating, an imaging lens or telescope, and the nonlinear material [1-3]. The disadvantage of this setup is the beam distortion caused by imaging errors limiting the useful pump spot size [14, 15] and, consequently, the THz energy. A more simple scheme was proposed by omitting the imaging optics and bringing the grating directly in contact with the crystal [14]. The advantage of such a contact-grating setup is that by eliminating imaging errors larger pumped areas can be efficiently used resulting in higher THz energies and better beam quality.

In this work we show that OR of femtosecond pulses by using the TPFP technique with a contact grating is a promising candidate to reach the high THz field strengths and pulse energies required by the applications mentioned above. The effect of varying the pump pulse duration and the crystal temperature (to minimize its THz absorption) will be investigated in detail by numerical calculations and optimal conditions will be given.

## 2. Theoretical model.

The applied theoretical model, described in detail in ref. [15], takes into account (i) the variation of the pump pulse duration (and therefore of the pump intensity) with the propagation distance due to material and angular dispersions [16, 17], (ii) the non-collinear propagation of pump and THz beams and (iii) the absorption in the THz range due to the complex dielectric function (determined by phonon resonances). All kinds of nonlinear effects but OR were neglected [18]. The wave equation with the nonlinear polarization was solved in the spectral domain (by using the slowly varying envelope approximation) [15, 18]. The temporal shape of the THz field was obtained by Fourier transformation.

In the calculations Gaussian pump pulses with a peak intensity of 40 GW/cm$^2$ were used, which is about half of the intensity where the onset of free-carrier absorption was experimentally observed in LN [19]. This is a practical choice, giving close-to-maximum THz yield, since free-carrier absorption limits the useful pump intensity [15]. For the pump wavelength 1064 nm was chosen (the exact choice does not significantly influence the results). The Fourier-limited pump pulse duration was varied between 50 fs and 1 ps. The pump pulse duration changes as it propagates through the LN crystal due to the angular dispersion of its spectral components, related to the pulse front tilt [16]. This effect is more pronounced in case of shorter pulse durations. In order to ensure the shortest possible pump pulse duration within the crystal, and hence the highest pump intensity, the pump pulses were assumed to be pre-chirped such that the Fourier-limited values are reached at the center of the crystal [15]. The constant peak intensity and pump beam diameter implied a corresponding increase of the pump pulse energy with increasing pulse duration. At each value of the pump pulse duration the electric field strength of the output THz radiation was maximized by choosing optimal crystal length and phase matching THz frequency. The phase matching frequency was iteratively fitted to the central frequency of the generated THz spectrum, which was depending on pump pulse duration. For the shortest pump pulse durations the crystal length was set to give maximal THz peak electric field strength at the crystal output. For longer pump pulses, where this choice would have resulted in crystal lengths exceeding 10 mm, the crystal length was set to 10 mm. Due to the large pulse-front-tilt angle (63º), and the associated walk-off effect, longer crystals are impractical even with cm-scale pump beam diameters. The crystal lengths used in the calculations are listed in table 1.

**Table 1.** The crystal lengths given in mm used in the calculations for various temperatures (*T*) and pump pulse durations ($\tau$).

| $\tau$ [fs] → <br> *T* [K] ↓ | 50 | 100 | 200 | 300 | 400 | 500 … 1000 |
|---|---|---|---|---|---|---|
| 300 | 0.95 | 2.2 | 6.5 | 7.0 | 8.5 | 10 |
| 100 | 1.5 | 4.4 | 9.0 | 10 | 10 | 10 |
| 10 | 1.5 | 5.0 | 10 | 10 | 10 | 10 |

In order to avoid photorefraction [20] and reduce THz absorption [21] stoichiometric LN (sLN) doped with 0.7 mol% Mg was assumed as the nonlinear medium. LN has significant absorption in the THz range at room temperature (figure 1). Since decreasing the temperature decreases absorption in the THz range [21], low temperature cases (100 K and 10 K) were also investigated in the calculations. Since the refractive index of sLN in the THz range is large ($n \approx 5.0$ at 1 THz), the Fresnel-loss at the output surface of the crystal is significant (about 45%). It was also taken into account in the calculations.

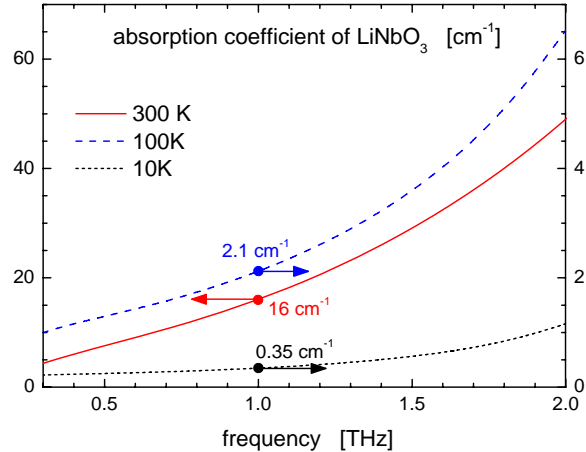

**Figure 1.** Frequency dependence of the absorption coefficient of LN in the THz range for different temperatures.

## 3. Results and discussion

### 3.1. Optimization for the electric field strength

The calculated THz spectra belonging to different pump pulse durations at 100 K are shown in figure 2(a). As it is obvious from figure 2(a), the peak spectral intensity increases and the position of the spectral intensity peak shifts towards lower frequencies with increasing pump pulse duration. This behavior can be observed for all temperatures, as shown for the central frequencies in figure 2(b). The central THz frequency varies between 1.5 and 0.27 THz for the investigated pump pulse duration range.

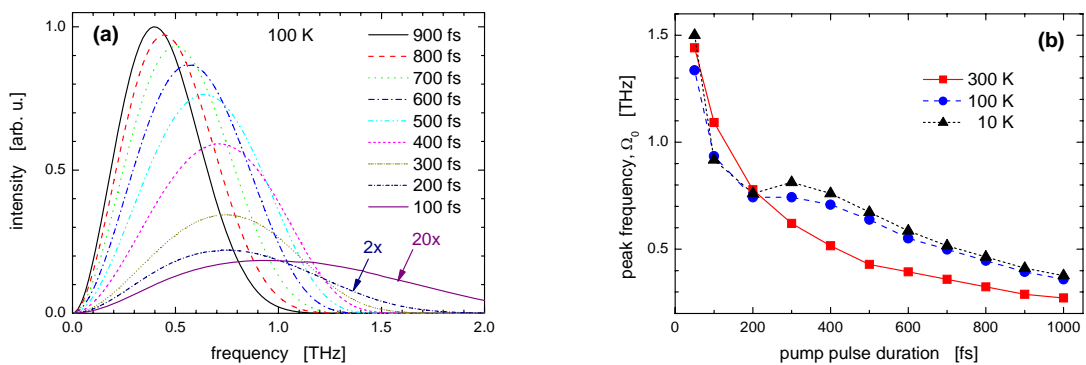

**Figure 2.** (a) THz spectra belonging to different pump pulse durations at 100 K temperature. (b) Frequency of the spectral peak of the THz pulses as function of the Fourier-limited pump pulse duration at different temperatures.

In order to calculate the peak electric field strength of the THz pulses their temporal shapes were calculated from the spectra by Fourier transformation. Examples are shown in figure 3(a). The calculated peak electric field strength of the THz pulses in air immediately after the output surface of the crystal is shown in figure 3(b) for different temperatures as a function of the Fourier-limited pump pulse duration ($\tau$). The encircled cross indicates the experimental conditions belonging to the values of 100 fs and 300 K. Similar parameters were used in many experiments pumped by amplified Ti:sapphire laser systems [1, 2, 22, 23]. For such experimental conditions our calculations give 240 kV/cm for the peak of the electric field strength (figure 3(b)). This exceeds by about a factor of two the value of 110 kV/cm obtained experimentally from the measured peak THz intensity and spot size [23]. Possible reasons for the difference are the shorter than 100 fs Fourier-limited pump pulse duration in experiments and imaging errors in the pulse-front-tilting setup [15]. The calculated value for the peak of the THz spectrum is 1.1 THz (figure 2(b)), which is similar to the values observed in experiments. This approximate agreement indicates that the present calculation method gives realistic predictions for the order of magnitude of the THz output in real experiments.

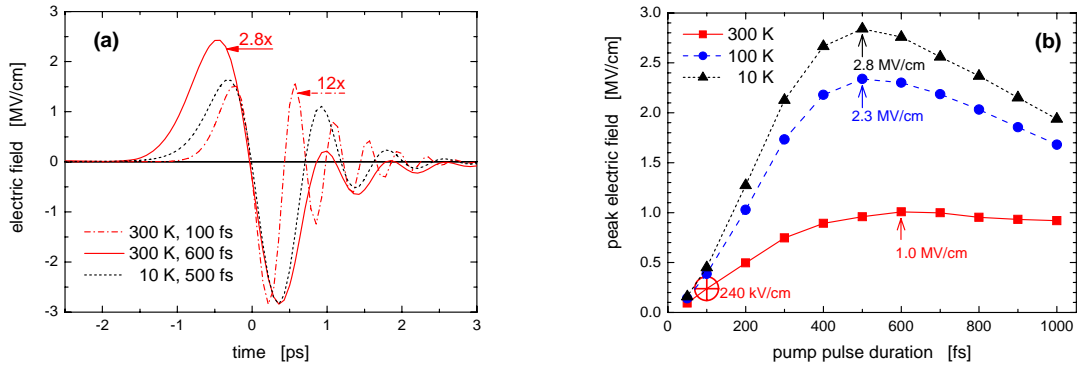

**Figure 3.** (a) Time dependence of the electric field strength of the generated THz pulses belonging to the optimal pump pulse durations for temperatures of 300 K and 10 K, as well as to 100 fs at 300 K. (b) Peak electric field strength of the THz pulses as function of the Fourier-limited pump pulse duration at different temperatures.

Increasing the pump pulse duration from the commonly used 100 fs results in significant changes in the THz peak electric field as well as in the frequency of the spectral peak. As shown in figure 3(b), by choosing the optimal pump pulse duration of 600 fs for room temperature the THz peak electric field strength can be increased by a factor of more than four, resulting in the extremely high value of 1.0 MV/cm at the output of the crystal. The position of the corresponding spectral peak is reduced to 0.4 THz (figure 2(b)). The reason of the increase in field strength is twofold: (i) The longer pump pulse causes a shift of the THz spectrum to lower frequencies, which results in reduced absorption within the crystal ($\alpha = 5.9$ cm$^{-1}$ at 0.4 THz instead of $\alpha = 18$ cm$^{-1}$ at 1.1 THz, see figure 1). (ii) The longer pump pulse also allows a longer THz generation length [15] (table 1).

Even higher electric field strength can be reached at lower temperatures. At temperatures of 100 and 10 K the maxima are located at 500 fs (figure 3(b)) with field strength values of 2.3 MV/cm and 2.8 MV/cm, respectively, corresponding to an enhancement of about one order of magnitude as compared to 300 K and 100 fs. The reason for this increase is clearly the reduced THz absorption at cryogenic temperatures (figure 1). In cases of optimal pump pulse duration (giving the highest peak electric fields) the central THz frequency has a value of 0.40 THz at 300 K, 0.64 THz at 100 K and 0.67 THz at 10 K temperatures as it is shown in figure 2(b).

Figure 3(a) shows the time-dependent electric field for the optimal pump pulse durations at 300 K and 10 K temperatures. For comparison, the THz pulse shape is also shown for 100 fs and 300 K closest to recent experimental parameters, as mentioned above. The amplitudes in figure 3(a) are scaled by different factors in order to obtain easier comparison of the pulse shapes with that belonging to the optimal case (500 fs and 10 K). It can be seen that a 12-fold increase in the peak electric field strength can be reached by cooling down the crystal to 10 K temperature and using optimal (500 fs) pump pulses.

*3.2. Contact-grating for extremely high energies and field strengths*

The calculated optical-to-THz conversion efficiencies are shown in figure 4(a) versus the pump pulse duration. The behavior is very similar to the electric field results shown in figure 3(b), but the maxima are slightly shifted towards shorter pulse durations. For example, the maxima are located at 400 fs for 10 and 100 K temperatures (instead of 500 fs, as in figure 3(b)). As expected, the difference between the efficiency curves belonging to different temperatures are even more pronounced than between the electric field curves of figure 3(b). We note that the influence of the THz field on the pump pulse, which can cause an increase of the THz generation efficiency [24, 25], can be significant in case of the very large efficiency values at optimal pump pulse durations at cryogenic temperatures. A more accurate numerical study requires to take these effects into account. The extremely high pump-to-THz energy efficiency values predicted by the calculations correspond to pump-to-THz photon conversion efficiency values exceeding 100%. We note that internal photon conversion efficiencies well above 100% are caused by cascade effects [26] and were indicated by recent experiments [1, 2].

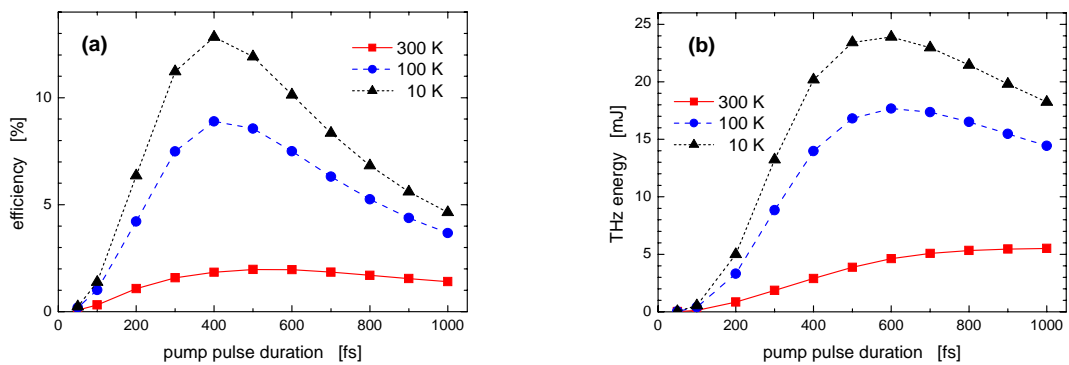

**Figure 4.** (a) Optical-to-THz conversion efficiency versus the Fourier-limited pump pulse duration at different temperatures. (b) THz pulse energy assuming a pump spot diameter of 5 cm versus the Fourier-limited pump pulse duration at different temperatures.

In order to fully exploit the scalability of the TPFP technique to extremely high THz pulse energies and field strengths it will be necessary to use a contact-grating setup [14] with a very large interaction area pumped by high-energy laser sources. We note that our preliminary experiments (with a prototype contact grating fabricated on ZnTe by laser ablation and pumped at 1.7 μm wavelength) indicate the working of the contact-grating scheme. Further work is in needed to optimize the grating characteristics. Tailoring the diffraction efficiency can be achieved by using binary gratings [27, 28].

Our calculations indicate that it is feasible to scale the THz pulse energy to the tens-of-mJ level by using the contact-grating scheme with LN. Figure 4(b) shows the calculated THz pulse energies obtained by using the efficiencies given in figure 4(a) and a pump beam diameter of 5 cm, which is a feasible beam size for use with the contact grating. As shown in figure 3(b), THz field strengths up to 2.8 MV/cm can be obtained by pumping a 10-mm thick LN crystal in a contact-grating setup with pulses of 500 fs duration, 40 GW/cm$^2$ peak intensity, and 5 cm beam diameter (about 200 mJ pump pulse energy) at 10 K temperature. According to figure 4(b), the corresponding output THz energy is 23 mJ. As compared to present experimental status this corresponds to an increase of two or three orders of magnitude in THz pulse energy, which will open up the field for various new applications.

A way to further increase the electric field strength of the THz output is the use of optimized focusing optics behind the crystal. For example, using an optical system consisting of two parabolic mirrors with focal lengths of 50 cm and 8 cm, and typical, commercially available aperture sizes, the electric field strength can be scaled to the 10 MV/cm level without any significant frequency cut-off. The contact-grating setup will allow to use large pump beam cross sections. Assuming an output THz beam diameter of 5 cm, and using a single parabolic mirror with focal length of 5 cm for focusing, a spot size of 0.38 mm can be reached in the focal plane at the central frequency. Even though in this setup the shape of the THz signal will be distorted because of diffraction [29], a peak electric field strength of about 100 MV/cm can be reached.

## 4. Conclusions

Numerical calculations were performed in order to maximize the electric field strength of THz pulses generated by tilted-pulse-front excitation in LN, motivated by various applications. It was shown that the pump pulse duration is a key experimental parameter in the THz generating process. According to the calculations the THz peak electric field strength can be increased by more than a factor of four to the MV/cm level directly at the crystal output by using 600 fs pump pulses instead of the commonly used 100 fs. The importance of the absorption of LN in the THz range was also discussed. The calculations predict about one order of magnitude increase in the THz peak electric field strength when the crystal is cooled to 10 K and 500 fs pump pulses are used, as compared to 100 fs pumping at room temperature. The electric field strength can easily be increased to the 10 MV/cm level by focusing. Using such optimized conditions in combination with the contact-grating technique will allow to generate THz pulses with peak electric field strength on the 100 MV/cm level and tens of mJ energy driven by efficient sub-joule class diode-pumped solid-state lasers. The predicted development of highly optimized sources for ultra-intense THz pulses will open up new applications in THz-assisted attosecond pulse generation, as well as in particle acceleration and manipulation by electromagnetic waves.

**Acknowledgements**
The authors acknowledge the fabrication of the grating structure on ZnTe to Cs. Vass and B. Hopp from Department of Optics and Quantum Electronics, University of Szeged. Financial support from Hungarian Scientific Research Fund (OTKA), grant numbers 76101 and 78262, and from "Science, Please! Research Team on Innovation" (SROP-4.2.2/08/1/2008-0011) is acknowledged.